\documentclass{svjour3}
\usepackage{hyperref}

\usepackage[utf8]{inputenc}
\usepackage[russian,english]{babel}
\usepackage{graphicx}
\usepackage{xcolor}

\definecolor{mybrown}{rgb}{0.5,0,0}
\definecolor{mylightgreen}{rgb}{0,0.83,0}
\definecolor{mygreen}{rgb}{0,0.5,0}

\usepackage{amssymb}
\usepackage{amsxtra}
\usepackage[mathscr]{eucal}
\usepackage{cite}

\makeatletter

\@addtoreset{equation}{section}
\makeatother
\journalname{CMAT}

\def\Re{\operatorname{Re}}
\def\I{\mathrm{i}}
\def\kappaT{T}
\def\F#1#2#3{#1_{{\mathscr F_{#3}^#2}}}

\def\AA{\bar a}

\def\varOmegaa{\bar\omega}
\def\breveC{{\mathscr C_0}}

\def\be#1{\begin{equation}\label{#1}}
\def\ee{\end{equation}}
\newcommand {\ba}[2]{\be{#1}\begin{array}{#2}}
\newcommand {\ea}{\end{array} \ee}
\def\eq#1{(\ref{#1})}

\renewcommand{\=}{\stackrel{\mbox{\scriptsize def}}{=}}

\def\({\left(}
\def\){\right)}
\def\av#1{\langle{#1}\rangle}

\let\w = \omega

\def\dt{\partial_t}

\def\LL{{\mathfrak L}}

\let\de = \delta

\def\L{{\mathcal L}}

\def\d{{\cal D}}

\let\ka=\kappa

\def\sign{\operatorname{sign}}

\def\qy{{y}}

\def\chio{\bar\chi_0}

\def\NEW#1{{\color{black}#1}}

\def\LL{\NEW{\omega_0^2\L}}

\def\bs#1{\boldsymbol{#1}}
\def\bb#1{\mathbf #1}
\def\d{\mathrm d}
\def\DDD{\mathrm d}
\def\bC{\bs{\mathscr C}}

\def\Ei{\operatorname{E}}
\def\semicolon{;}

\begin{document}
\selectlanguage{english}
\title{Steady-state kinetic temperature distribution in a
two-dimensional square harmonic scalar lattice lying in a viscous environment
and subjected to a point heat source\thanks{
This work is supported by Russian Science Foundation (Grant No. 18-11-00201).}}
\author{Serge N.~Gavrilov \and Anton M.~Krivtsov}

\titlerunning{Steady-state kinetic temperature distribution}
\institute{
S.N.~Gavrilov \at
Institute for Problems in Mechanical Engineering RAS, V.O., Bolshoy pr.~61,
St.~Petersburg, 199178, Russia \\
\email{serge@pdmi.ras.ru}           
\and
S.N.~Gavrilov \at
Peter the Great St.~Petersburg Polytechnic University (SPbPU),
Polytechnicheskaya str.~29, St.Petersburg, 195251, Russia
\and
A.M.~Krivtsov \at
Peter the Great St.~Petersburg Polytechnic University (SPbPU),
Polytechnicheskaya str.~29, St.Petersburg, 195251, Russia\\
\email{akrivtsov@bk.ru}
\and
A.M.~Krivtsov \at
Institute for Problems in Mechanical Engineering RAS, V.O., Bolshoy pr.~61,
St.~Petersburg, 199178, Russia}

\selectlanguage{english}
\maketitle

\begin{abstract}	
We consider heat transfer in an infinite two-dimensional square harmonic scalar lattice
lying in a viscous environment and subjected to a heat source.  The
basic equations for the particles of the lattice are stated in the form of a
system of stochastic ordinary differential equations. We perform a continualization
procedure and derive an infinite system of linear partial differential equations
for covariance variables.  The most important results of the paper are the
deterministic
differential-difference equation describing non-stationary heat propagation in
the lattice and the analytical formula in the integral form for its 
steady-state solution describing kinetic temperature distribution
caused by a point
heat source of a constant intensity.  The comparison between numerical solution
of stochastic equations and obtained analytical solution 
demonstrates a very good agreement
everywhere except for the main diagonals of the lattice (with respect to the point
source position), where the analytical solution is singular. 
\keywords{ballistic heat transfer \and 2D harmonic scalar lattice \and kinetic
temperature}
\end{abstract}

\let\d=\DDD
\def\F#1#2#3{#1_{{F}}}

\section{Introduction}

At the macroscale, Fourier's law of heat
conduction is widely and successfully used to describe heat transfer
processes.
However, recent experimental
observations demonstrate that Fourier's law is violated at the microscale and
nanoscale, in particular, in
low-dimensional nanostructures
\cite{chang2008breakdown,%
xu2014length,%
hsiao2015micron,%
cahill2003nanoscale,%
liu2012anomalous,%
lepri2016thermal8}, where the ballistic heat transfer is
realized. {The anomalous heat transfer also may be related with the
spontaneous emergence of long-range correlations; the latter is typical
for momentum-conserving systems \cite{lepri2003thermal,spohn2016fluctuating}.}
The simplest theoretical approach to describe the ballistic heat propagation
is to use harmonic lattice models. In some cases such
models allow one to obtain the analytical description of thermomechanical
processes in solids
\cite{hoover2015simulation,daly2002molecular,krivtsov2003nonlinear,berinskii2016elastic,kuzkin2016lattice
,
berinskii2015linear,berinskii2016hyperboloid}.
In the literature, the problems concerning heat transfer in harmonic lattices are mostly
considered in the stationary formulation 
\cite{%
bonetto2000mathematical,%
lepri2003thermal,%
0295-5075-43-3-271,%
dhar2008heat,%
rieder1967properties,%
allen1969energy,%
nakazawa1970lattice,%
lee2005heat,%
kundu2010heat,%
lepri2016thermal2,%
bernardin2012harmonic,%
freitas2014analytic,%
freitas2014erratum,%
hoover2013hamiltonian,%
lukkarinen2016harmonic},
the non-stationary heat propagation is discussed in
\cite{le2008molecular,%
gendelman2012nonstationary,%
tsai1976molecular,%
ladd1986lattice,%
volz1996transient,%
daly2002molecular,%
gendelman2010nonstationary,%
guzev2018
}.

In previous studies
\cite{krivtsov2014energy,krivtsov2015heat,krivtsov-da70},
a new approach was suggested which allows one to solve
analytically non-stationary thermal problems for an infinite one-dimensional harmonic
crystal --- an infinite ordered chain of identical material particles,
interacting via linear (harmonic) forces. In particular, a heat transfer
equation was obtained that differs from the extended heat transfer equations
suggested earlier \cite{chandrasekharalah1986thermoelasticity,tzou2014macro,cattaneo1958forme,vernotte1958paradoxes};
however, it is in an excellent agreement with
molecular dynamics simulations and previous analytical estimates 
\cite{gendelman2012nonstationary}. 
The properties of the solutions describing heat transfer in a one-dimensional
harmonic crystal were discussed in
\cite{%
sokolov2017localized,
krivtsov2018one,
PhysRevE.99.042107}.
Later
this approach was generalized 
\cite{%
babenkov2016energy,
krivtsov2014energy,
krivtsov2015heat,
kuzkin2017analytical,
kuzkin2017high,
Kuzkin-Krivtsov-accepted,
murachev2018thermal}
to a number of systems, namely, to an 
infinite one-dimensional
crystal on an elastic substrate 
\cite{babenkov2016energy}, 
to an infinite one-dimensional diatomic harmonic crystal
\cite{podolskaya2018anomalous},
to a finite 
one-dimensional
crystal \cite{murachev2018thermal},
and to two and three-dimensional
infinite harmonic lattices
\cite{kuzkin2017analytical,Kuzkin-Krivtsov-accepted,kuzkin2017high}.
In the most of the above mentioned papers 
\cite{
babenkov2016energy,
krivtsov2014energy,
krivtsov2015heat,
kuzkin2017analytical,
kuzkin2017high,
Kuzkin-Krivtsov-accepted,
murachev2018thermal,
sokolov2017localized}
only
isolated systems were considered.

In recent paper 
\cite{gavrilov2018heat}
an infinite one-dimensional harmonic crystal that can exchange energy
with its surroundings was considered.
It was assumed that the crystal lies in a viscous
environment (a gas or a liquid, e.g., the air)
which causes an additional dissipative term in
the equations of stochastic dynamics for the particles. Additionally, 
sources of heat supply were taken into account as an additional noise term in
equations of motions. Unsteady heat conduction
regimes as well as the steady-state kinetic temperature distribution caused by
a point heat source of a constant intensity were investigated.

\begin{figure}[htbp]	
\centering\includegraphics[width=0.8\textwidth]{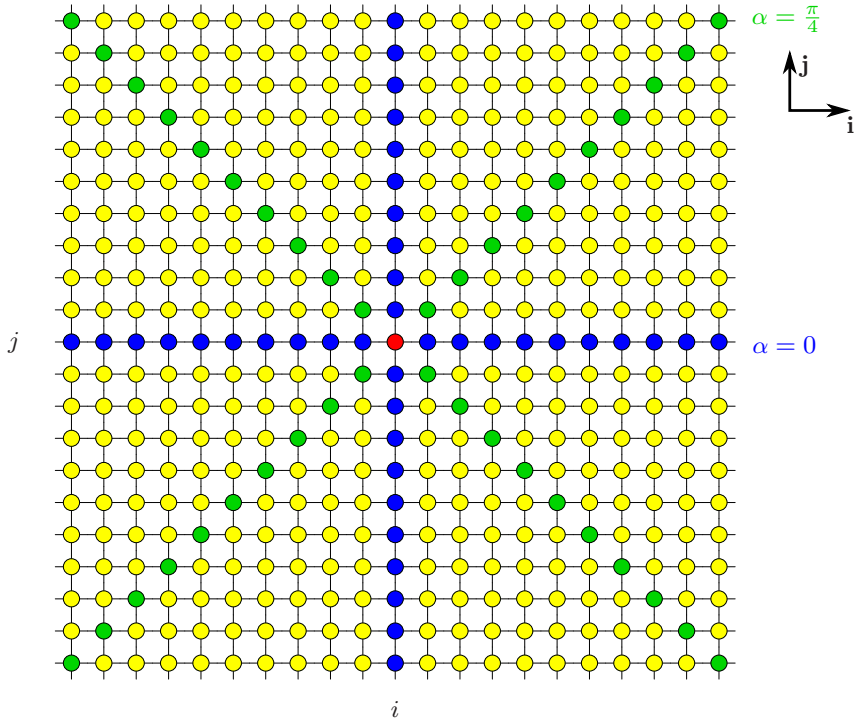}
\caption{Schematic of the lattice subjected to a point heat source:
the particle, where the point heat source is
applied, is shown by the red color; two orthogonal rows of the lattice that
contain the heat source are shown by the blue color; the main diagonals of the
lattice (with respect to the point heat 
source position) are shown by the green color}
\label{lattice.eps}
\end{figure}

In the present paper we generalize to the two-dimensional case the results 
obtained in paper \cite{gavrilov2018heat}.
We use the model of an infinite
two-dimensional 
harmonic square scalar\footnote{This means that the displacements are scalar quantities
\cite{mielke2006macroscopic,harris2008energy,savin2016normal,nishiguchi1992thermal,Kuzkin-Krivtsov-accepted}}
lattice performing transverse oscillations. 
The
schematic of the system is shown in Figure~\ref{lattice.eps}.
The aim of the paper is to develop an equation describing non-stationary
heat propagation in the lattice and use this equation to obtain
analytical formulas describing the steady-state kinetic temperature
distribution caused by a point heat source of a constant intensity. The
differences in methodology between the present paper and
\cite{gavrilov2018heat} are:
\begin{itemize}	
\item
When deriving the differential-difference equation describing non-stationary
propagation in the lattice, we avoid to use identities of the calculus of
finite differences and instead inverse the corresponding finite difference
operators, when necessary, in the Fourier domain. 
This significantly
simplifies the formal procedure. Accordingly, we do not introduce into consideration
the quantities, which we call ``non-local temperatures'' in
\cite{gavrilov2018heat} (Section~4) and deal with
covariance variables only.
\item
Since the two-dimensional case is more complicated, the steady-state solution
for the kinetic temperature is obtained as a solution of
the stationary equations for covariances of the particle velocities (in
\cite{gavrilov2018heat} the steady-state solution  was found as a limiting
case for the
corresponding unsteady solution). 
\item
The
investigation of unsteady heat conduction regimes is beyond the scope of this
paper.
\end{itemize}

The paper is organized as follows. In Section~\ref{2d-Sec-formulation}, we
consider the formulation of the problem. In Section~\ref{2d-SSec-notation},
some general notation is introduced. In Section~\ref{2d-SSec-stochastic}, we state the basic
equations for the lattice particles in the form of a system of 
stochastic differential equations. In Section~\ref{2d-SSec-covariance}, we
introduce and deal with infinite set of covariance variables. These  
are the mutual covariances of all the particle velocities and all the
displacements for all pairs of particles. We use the It\^o lemma to derive
(see \cite{gavrilov2018heat}, Appendix~A) an infinite deterministic system of ordinary
differential equations which follows from the equations of stochastic dynamics.
This system can be transformed into an infinite system of differential-difference
equations involving only the covariances for {the particle velocities}.
In Section~\ref{2d-Sec-cont}, we introduce a vectorial continuous spatial variable 
and write the finite difference operators involved in the equation for
covariances as compositions of finite difference operators and operators of
differentiation.
In Section~\ref{2d-Sec-slow}, we perform an asymptotic uncoupling
of the equation for covariances. Provided that the introduced continuous spatial
variable can characterize the behavior of the lattice, one can
distinguish between slow motions, which are related to the heat propagation, and
{vanishing fast motions
\cite{krivtsov2014energy,babenkov2016energy,kuzkin2017high,Kuzkin-Krivtsov-accepted,kuzkin2017analytical,kuzkin2019thermal}}, which are not considered in the paper. 
Slow motions
can be described by a coupled infinite system of second-order hyperbolic partial
differential equations  for the continualized covariances of velocities. 
The kinetic temperature is introduced as a quantity proportional to the
variance of 
the particle velocities.
In Section~\ref{2d-stationary} we 
obtain an analytical expression in the integral form for the steady-state 
solution describing the kinetic temperature distribution caused by 
a point source of a constant 
intensity (the corresponding calculations can be found in
Appendices 
\ref{2d-app-A}
\&
\ref{2d-app-B}). 
In Section~\ref{2d-sec-numerics}, we present the
results of the numerical solution of the initial value problem for the system
of stochastic differential equations and compare them with the obtained analytical
solution.
In the conclusion (Section~\ref{2d-Sec-conclusion}), we discuss the basic
results of the paper.

\section{Mathematical formulation}
\label{2d-Sec-formulation}
\subsection{Notation}
\label{2d-SSec-notation}
In the paper, we use the following general notation:
\begin{description}	
\item[$t$] is the time;
\item[$H(\cdot)$] is the Heaviside function;
\item[$\delta(\cdot)$] is the Dirac delta function in a two-dimensional space;
\item[$\delta_1(\cdot)$] is the Dirac delta function in a one-dimensional space;
\item[$\langle\cdot\rangle$] is the expected value for a random quantity;
\item[$\de_{p;q}$] is
the Kronecker delta ($\de_{p;q}= 1$ if  $p=q$, and $\de_{p;q} = 0$
otherwise);
\item[$\mathbb Z$] is
the set of all integers;
\item[$\mathbb R$] is
the set of all real numbers. 
\end{description}	

\subsection{Stochastic lattice dynamics}
\label{2d-SSec-stochastic}

Consider the following infinite system of stochastic ordinary differential equations
\cite{kloeden1999,stepanov2013stochastic}:
\begin{gather}
\d v_{i,j} = F_{i,j}\, \d t + b_{i,j}\, \d W_{i,j},
 \qquad
\d u_{i,j} = v_{i,j}\, \d t,
\label{2D-1}
\end{gather}
where
\begin{gather}
F_{i,j}=\LL_{i,j} u_{i,j} - \eta v_{i,j},
\label{2D-F_i}
\\
\d W_{i,j}= \rho_{i,j} \sqrt{\d t}
\label{2D-Winer},\\
\w_0 \= \sqrt{C/m},
\end{gather}
{Here $i,j\in\mathbb Z$ are 
arbitrary integers which
describes the position of a particle in the square lattice; 
the stochastic processes $u_{i,j}(t)$ and $v_{i,j}(t)$ are the displacement and
the particle velocity, respectively;
$F_{i,j}$ is the force on the
particle;  
$W_{i,j}$ are uncorrelated Wiener processes \cite{kloeden1999,stepanov2013stochastic};
$b_{i,j}(t)$ is the intensity of the random external excitation;
$\eta$ is the specific viscosity for the environment;
$C$ is the bond stiffness; $m$ is the mass of a particle;
$\L_{i,j}$ is the linear finite
difference operator describing an infinite two-dimensional stretched
square scalar lattice with nearest-neighbor interactions performing out-of-plane
vibrations:
\begin{gather}
\L_i u_{i,j}=u_{i+1,j} - 2u_{i,j} + u_{i-1,j},
\label{2D-Li}
\\
\L_{i,j} u_{i,j}=
\L_{i} u_{i,j}+
\L_{j} u_{i,j}
\label{2D-Lij}
\end{gather}
}

The normal random variables $\rho_{i,j}$ are such that 
\begin{equation}
    \av{\rho_{i,j}} = 0, \qquad
    \av{\rho_{i,j}\rho_{k,l}} = 
    \delta_{i;k}\delta_{j;l}
    \label{2D-82}
\end{equation}
and they are assumed to be independent of $u_i$ and $v_i$.
The initial conditions are zero: for all $i$,~$j$
\begin{equation}
u_{i,j}(0)=0,\qquad v_{i,j}(0)=0.
\label{2D-ic-stochastic}
\end{equation}

In the case $b_{i,j}\equiv b$, equations \eqref{2D-1} are the Langevin equations
\cite{langevin1908theorie,lemons1997paul} for a
two-dimensional harmonic stretched
square lattice
surrounded by a viscous environment (e.g., a gas or a liquid). Assuming that
$b_{i,j}$ may depend on ${i,j}$, we introduce a natural generalization of the Langevin
equation which allows one to describe the possibility of an external heat
excitation (e.g.,\ laser excitation) \cite{gavrilov2018heat}. 
This external excitation is assumed to be localized in space:
\begin{equation}
b_{i,j}\equiv0
\quad \text{if} \quad\max\{i,j\}>R
\label{2d-localized}
\end{equation}
for some $R>0$,
{and much more intensive than the stochastic influence caused by a non-zero
temperature of the environment.} Therefore,
we neglect 
in 
\eqref{2D-1}
the stochastic term, which variance does not depend on $i$ and $j$, that
models the influence from the environment.\footnote{To consider the
full problem with a non-zero temperature of the environment we need to take
into account the processes of thermal equilibration (the fast motions) being coupled
with the heat propagation. The preliminary calculations show that the presence of
the viscous environment makes the problem to be much more complicated in
comparison with the conservative case considered, e.g., in 
\cite{Kuzkin-Krivtsov-accepted}. 
At the same time, in the framework of this general problem,
the processes of heat transfer can be easily separated, and due to linearity
of the general problem, they described by exactly the same equations as
ones formulated in this paper.}
Note that zero initial
conditions 
\eqref{2D-ic-stochastic} and requirement 
\eqref{2d-localized} guarantee that there are no sources at the infinity,
thus we do not need additional boundary conditions at the infinity.

\subsection{The dynamics of covariances}
\label{2d-SSec-covariance}

According to \eqref{2D-F_i}, 
$F_{i,j}$ are linear functions of $u_{i,j},\ v_{i,j}$.
Taking
this fact into account together with \eqref{2D-82} and \eqref{2D-ic-stochastic}, 
we see that for all $t$
\begin{equation}
\av{u_{i,j}}=0,\qquad \av{v_{i,j}}=0.
\end{equation}
Following \cite{krivtsov2016ioffe,gavrilov2018heat}, consider the infinite sets of
covariance variables\footnote{We use commas to separate the subscripts related
with one particle, and semicolon to separate the subscripts related with
different particles.}
\begin{equation}
    \xi_{p,q{\semicolon}r,s} \= \av{u_{p,q}\, u_{r,s}}, \qquad
    \nu_{p,q{\semicolon}r,s} \= \av{u_{p,q}\, v_{r,s}}, \qquad
    \ka_{p,q{\semicolon}r,s} \= \av{v_{p,q}\, v_{r,s}},
\label{2D-4}
\end{equation}
and the {quantities}
\begin{equation}
\beta_{p,q{\semicolon}r,s} \= \delta_{p;r}\delta_{q;s}\,b_{p,q} b_{r,s}.
\label{2D-beta-pq}
\end{equation}
In the last formula, we take into account the second formula of \eqref{2D-82}.
Thus the variables 
$\xi_{p,q{\semicolon}r,s}$,
$\nu_{p,q{\semicolon}r,s}$,
$\ka_{p,q{\semicolon}r,s}$
are defined
for any pair of lattice particles.
For simplicity, in what follows we drop all the
subscripts, i.e., $\xi\=\xi_{p,q{\semicolon}r,s}$ etc. By definition, we also put
$\xi^\top\=\xi_{r,s{\semicolon}p,q}$ etc. 
Now we differentiate the variables 
\eqref{2D-4} with respect to time
taking into account the equations of motion \eqref{2D-1}. This yields
the following closed system of differential equations for
the covariances (see \cite{gavrilov2018heat}, Appendix~A):
\begin{equation}	
\begin{gathered}	
    \dt\xi = \nu + \nu^\top,\\
    \dt\nu + \eta\nu = \LL_{r,s}\,\xi + \ka,\\
    \dt\ka + 2\eta\ka = \LL_{p,q}\nu + \LL_{r,s}\nu^\top + \beta,
\end{gathered}
\label{2D-5}
\end{equation}
where $\dt$ is the operator of differentiation with respect to time;
{{$\L_{p,q}$ and $\L_{r,s}$ are the linear difference operators defined by 
\eqref{2D-Lij} that act on  
$\xi_{p,q{\semicolon}r,s}$,
$\nu_{p,q{\semicolon}r,s}$,
$\ka_{p,q{\semicolon}r,s}$,
$\beta_{p,q{\semicolon}r,s}$
with respect to the first pair of subscripts $p,\ q$ and the second one $r,\
s$, respectively.}}
Now we introduce the symmetric and antisymmetric difference operators
\begin{equation}
2\L^{\mathrm S} \= \L_{p,q}+\L_{r,s},\qquad
2\L^{\mathrm A} \= \L_{p,q}-\L_{r,s},
\end{equation}
and the symmetric and antisymmetric parts of the variable $\nu$:
\begin{equation}
    2\nu^{\mathrm S} \= \nu+\nu^\top, \qquad
    2\nu^{\mathrm A} \= \nu-\nu^\top.
\label{2D-9}
\end{equation}
Note that $\xi$ and $\kappa$ are symmetric variables.
Now equations \eq{2D-5} can be rewritten as follows: 
\begin{gather}
\dt\xi = 2\nu^{\mathrm S},
\qquad(\dt + 2\eta)\ka =  2\LL^{\mathrm S}\nu^{\mathrm S} 
+ 2\LL^{\mathrm A}\nu^{\mathrm A} + \beta,
\label{2D-11}
\\
(\dt + \eta)\nu^{\mathrm A} = -\LL^{\mathrm A}\xi,
\qquad(\dt + \eta)\nu^{\mathrm S} = \LL^{\mathrm S}\xi + \ka.
\label{2D-11-post}
\end{gather}
This system of equations can be reduced 
(see \cite{gavrilov2018heat})
to one equation of the fourth order in
time for covariances of the particle velocities $\kappa$:
\begin{equation}
\((\dt + \eta)^2(\dt^2 + 2\eta\dt - 4
\LL^{\mathrm S}) 
+ 4
(\LL^{\mathrm A})^2\)\ka =
(\dt+\eta)(\dt^2 + \eta\dt - 2
\LL^{\mathrm S})\beta.
\label{2D-30}
\end{equation}
In what follows, we deal with Eq.~\eqref{2D-30}. 
This equation can describe a wide class of systems. For example, in the simplest particular
case $\mathscr L_{i,j}u_{i,j}=-u_{i,j}$ it can be used to obtain the classical
results 
\cite{wang1945theory,stepanov2013stochastic}
concerning time evolution of the
variance of the velocity for a single stochastic oscillator with additive
noise.

According to 
Eqs.~\eqref{2D-ic-stochastic}, \eqref{2D-4}, we supplement Eq.~\eqref{2D-30}
with zero initial conditions. We state these conditions in the following form, 
which is conventional for distributions (or generalized functions)~\cite{Vladimirov1971}:
\begin{equation}
\kappa\big|_{t<0}\equiv0.
\label{2D-ic<0-pre}
\end{equation}
To take into account non-zero classical initial conditions, one needs to add 
the corresponding singular terms (in the form of a linear combination of 
$\delta(t)$ and its derivatives) to the right-hand sides of the corresponding equations 
\cite{Vladimirov1971}.

{Let us note that  equation \eqref{2D-30} is
a deterministic equation}. What is also important is that \eqref{2D-30} is a closed
equation. Thus, the thermal processes do not depend on any property of the
cumulative distribution functions for the displacements and the particle
velocities other than the covariance variables used above.

\section{Continualization of the finite difference operators}
\label{2d-Sec-cont}

Following \cite{krivtsov2015heat,krivtsov2016ioffe,gavrilov2018heat}, we 
introduce the discrete spatial variables
\begin{equation}
k\= p+r,\qquad 
l\= q+s
\label{2D-k-def}
\end{equation}
and the discrete correlational variables
\begin{equation}
m\=r-p,\qquad
n\=s-q.
\label{2D-n}
\end{equation}
instead of discrete variables $p$, $q$, $r$, $s$.
We can also formally introduce the continuous vectorial spatial variable 
\begin{equation}
\bb x\=\frac{a}2(k\bb i+l\bb j),
\label{2D-x-def}
\end{equation}
where $a$ is {the lattice constant} (the distance between neighboring
particles along a row), $\mathbf i$ and $\mathbf j$ are orthogonal unit vectors ({the directions of the principal
axes of the lattice}, see Figure~\ref{lattice.eps}). 
In what follows, we represent $\bb x$ also in the following form:
\begin{equation}
\bb x\= x_1\bb i+x_2\bb j,
\label{r-def}
\end{equation}
where 
$x_1,\ x_2$ are the spatial coordinates in the lattice.

To perform the continualization, we assume that the lattice constant is a
small quantity and introduce a dimensionless formal small parameter
$\epsilon$ in the following way:
\begin{gather}
a=\epsilon \bar a,
\label{2D-e-a}
\end{gather}
where $\AA=O(1)$. 
To preserve the speed of sound in the lattice
\begin{equation}
c\=a\w_0
\end{equation}
as a quantity of order $O(1)$, we additionally assume that
\begin{gather}
\omega_0=\epsilon^{-1}\varOmegaa_0,
\label{2D-epsilon}
\end{gather}
where $\varOmegaa_0=O(1)$. Thus
$c=\bar a\varOmegaa_0=O(1)$.
The basic assumption that allows one to perform the continualization is
that any {quantity} $\zeta_{p,q{\semicolon}r,s}$ defined by \eqref{2D-4} 
or \eqref{2D-beta-pq} 
can be calculated as a value of a smooth function
$\hat\zeta_{m,n}(\mathbf x)$ of the continuous spatial slowly varying coordinate
$
\mathbf x=\tfrac12\epsilon\AA (k\bb i+l\bb j)
$
and the discrete correlational
variables $m$ and $n$:
\begin{equation}
\hat\zeta_{m,n}(\bb x)=
\zeta_{p,q{\semicolon}r,s}.
\label{2D-Z_n}
\end{equation}
In accordance with Eqs.~\eqref{2D-k-def}, \eqref{2D-n},
one has
\begin{equation}
\begin{aligned}	
&\L_{p,q}\zeta_{p,q{\semicolon}r,s}=
&&\hat\zeta_{m-1,n}
\left(\bb x+\tfrac a2\bb i\right)
+\hat\zeta_{m+1,n}
\left(\bb x-\tfrac a2\bb i\right)
\\
&&+\,&
\hat\zeta_{m,n-1}
\left(\bb x+\tfrac a2\bb j\right)
+\hat\zeta_{m,n+1}
\left(\bb x-\tfrac a2\bb j\right)
-
4\hat\zeta_{m,n}(\bb x)
,
\\
&\L_{r,s}\zeta_{p,q{\semicolon}r,s}=
&&\hat\zeta_{m+1,n}
\left(\bb x+\tfrac a2\bb i\right)
+\hat\zeta_{m-1,n}
\left(\bb x-\tfrac a2\bb i\right)
\\
&&+\,&
\hat\zeta_{m,n+1}
\left(\bb x+\tfrac a2\bb j\right)
+\hat\zeta_{m,n-1}
\left(\bb x-\tfrac a2\bb j\right)
-
4\hat\zeta_{m,n}(\bb x)
.
\end{aligned}
\end{equation}
Applying the Taylor theorem to these formulas yields
\begin{equation}
\begin{aligned}	
&\L_{p,q}\zeta_{p,q{\semicolon}r,s}=
\L_{m,n}\hat\zeta_{m,n}
&+\,&
\tfrac a2\,\partial_{x_1}(\hat\zeta_{m-1,n}-\hat\zeta_{m+1,n})
\\
&&+\,&
\tfrac a2\,\partial_{x_2}(\hat\zeta_{m,n-1}-\hat\zeta_{m,n+1})
+o(\epsilon^2),
\\
&\L_{r,s}\zeta_{p,q{\semicolon}r,s}=
\L_{m,n}\hat\zeta_{m,n}
&+\,&
\tfrac a2\,\partial_{x_1}(\hat\zeta_{m+1,n}-\hat\zeta_{m-1,n})
\\
&&+\,&
\tfrac a2\,\partial_{x_2}(\hat\zeta_{m,n+1}-\hat\zeta_{m,n-1})
+o(\epsilon^2).
\end{aligned}
\label{2D-LpLq-cont}
\end{equation}
An alternative way of continualization can be realized by letting
the number of particles diverge, rather than invoking an increasingly
small separation \cite{lepri2010nonequilibrium}.
Despite the algorithmic difference, these approaches apparently lead to the same result,
although application of the first approach to an infinite system seems to be
more straightforward.

Now we perform the continualization of the operators 
$\L^{\mathrm S},\ \L^{\mathrm A}$.
Using
\eqref{2D-LpLq-cont}, we obtain
\begin{gather}	
\L^{\mathrm S}
=\L_{m,n}+O(\epsilon^2),
\label{2D-L-S}
\\
\L^{\mathrm A}=
-\tfrac a2\,\mathscr D_{m}\partial_{x_1}
-\tfrac a2\,\mathscr D_{n}\partial_{x_2}+
O(\epsilon^2).
\label{2D-LA-pre-pre}
\end{gather}
where 
\begin{equation}
\mathscr D_n f_n=f_{n+1}-f_{n-1}.
\end{equation}

\section{Slow motions}
\label{2d-Sec-slow}

Taking into account 
assumption \eqref{2D-epsilon},
Eq.~\eqref{2D-30} can be rewritten in the following form:
\begin{multline}
\bigg((\dt + \eta)^2\big(\epsilon^2(\dt^2 + 2\eta\dt) -
4\varOmegaa_0^2\L^{\mathrm S}\big) 
+ 4\frac{\varOmegaa_0^4}{\epsilon^2}(\L^{\mathrm A})^2\bigg)\ka \\=
(\dt+\eta)\big(\epsilon^2(\dt^2 + \eta\dt) - 2\varOmegaa_0^2\L^{\mathrm S}\big)\beta.
\label{2D-slow-epsilon}
\end{multline}
Equation~\eqref{2D-slow-epsilon} is a differential equation whose highest
derivative with respect to $t$ is multiplied by a small parameter. Therefore,
one can expect the existence of two types of solutions, namely, solutions slowly varying in time
and fast varying in time 
\cite{nayfeh2008perturbation,kevorkian2012multiple}. 
{The presence of fast and slow motions is a standard property of
statistical systems. Fast motions are oscillations
of temperature caused by equilibration of kinetic and potential energies. Slow
motions are related with macroscopic heat propagation.}

Considering slow
motions, we assume that 
\begin{equation}
\epsilon^2(\dt^2+2\eta\dt)\kappa\ll\varOmegaa_0^2\L^{\mathrm S}\kappa,
\qquad
\epsilon^2(\dt^2+\eta\dt)\beta\ll\varOmegaa_0^2\L^{\mathrm S}\beta.
\label{2D-slow-condition}
\end{equation}
Vanishing solutions 
\cite{krivtsov2014energy,babenkov2016energy,kuzkin2017high,Kuzkin-Krivtsov-accepted,kuzkin2017analytical}
that characterize fast motions, which do not satisfy 
\eqref{2D-slow-condition}, are not considered in this paper.

Now, taking into account Eqs.~\eqref{2D-L-S}, \eqref{2D-LA-pre-pre},
we drop the higher order terms and rewrite equation 
\eqref{2D-slow-epsilon} in the form of an equation for slow motions:
\begin{equation}
\Big((\dt + \eta)^2\L_{m\,n} - \tfrac {c^2}4
\big(\mathscr D_{m}\partial_{x_1}+
\mathscr D_n\partial_{x_2}\big)^2
\Big)
    \hat\ka_{m,n} =
    \frac12(\dt+\eta)\L_{m,n}\hat\beta_{m,n},
\label{2D-57}
\end{equation}
where quantity $\hat\beta_{m,n}(\bb x)$ 
is introduced in accordance with Eq.~\eqref{2D-Z_n}.
Note that 
according to 
\eqref{2D-beta-pq}, 
$\hat\beta_{m,n}=0$ if $m\neq0$ or $n\neq0$, thus 
\begin{equation}
\hat\beta_{m,n}=\hat\beta_{0,0}(\mathbf x)\delta_{m;0}\delta_{n;0}.
\end{equation}
We identify the following quantities
depending on the continuous variable $\bb x$
\begin{gather}
     T\equiv\theta_{0,0}(x,t) \= mk_B^{-1} \hat\ka_{0,0}(\mathbf x,t),
\label{2D-temp-def}     
\\
     \chi(\bb x,t) \= \tfrac12m k_B^{-1} \hat\beta_{0,0}(\mathbf x,t).
\label{2D-chi-def}
\end{gather}
as the kinetic temperature
and the heat supply intensity, respectively. 
Here $k_B$ is the Boltzmann constant.
We also introduce the following notation:
\begin{equation}
\theta_{m,n}(x,t) \= mk_B^{-1} \hat\ka_{m,n}(\mathbf x,t).
\label{theta_m_n}
\end{equation}
Now Eq.~\eqref{2D-57} can be rewritten as the following infinite system of PDE
describing non-stationary
heat propagation in the lattice:%
\footnote{These equations have a little bit different structure than the
corresponding equations
obtained in previous paper
(\cite{gavrilov2018heat}, Eq.~(4.3)), since 
the quantities $\theta_{m,n}(x,t)$ ($mn\neq0$) differ from the quantities,
which we call ``non-local temperatures''
in \cite{gavrilov2018heat} (see Eq.~(4.7)).}
\begin{equation}
    \Big((\dt + \eta)^2\L_{m,n} - \tfrac {c^2}4
\big(\mathscr D_{m}\partial_{x_1}+
\mathscr D_n\partial_{x_2}\big)^2   
\Big)\,
    \theta_{m,n} = (\partial_t\chi+\eta\chi)\L_{m,n}\delta_{m;0}\delta_{n;0}.
\label{2D-maineq}
\end{equation}
The initial conditions that corresponds to 
Eq.~\eqref{2D-ic<0-pre} are
\begin{equation}
\theta_{m,n}(\bb x,t)\big|_{t<0}\equiv0.
\label{ic<0}
\end{equation}
In the particular case where $\eta=0,\ \chi\propto\delta(t)$,
system \eqref{2D-maineq} was obtained and investigated in
\cite{Kuzkin-Krivtsov-accepted}. In the latter case the derivation of the
corresponding system of PDE is much simpler,
since it can be based on consideration of system of ordinary differential
equations with random initial conditions instead of system of stochastic
equations \eqref{2D-1}. 

Note that equations \eqref{2D-maineq} for slow motions involve only the product
$c=\omega_0a$ and do not involve the quantities $\omega_0$ and $a$ separately, 
so they do not involve $\epsilon$. Provided that the initial conditions
also do not involve $\epsilon$, the solution of the corresponding initial
value problem and all its derivatives are quantities of order $O(1)$.  The
rate of vanishing for fast motions depends on $\epsilon$: the smaller
$\epsilon$, the higher the rate. Thus, for sufficiently small $\epsilon$,  exact
solutions of Eq.~\eqref{2D-30}
quickly transform into slow motions.

\section{The steady-state solution of equation for slow motions}
\label{2d-stationary}
In what follows, we look for the steady-state solution of initial value
problem for system of partial differential equations \eqref{2D-maineq}
where the heat supply is given in the form of a point source of a constant
intensity. Accordingly, we take
\begin{gather}
\chi={\bar\chi_0}H(t)\delta(\bb x),
\label{2d-chi-H}
\end{gather}
where $\bar\chi_0=\mathrm{const}$.  Looking for the steady-state solution, we 
substitute $1(t)$ instead of $H(t)$ into Eq.~\eqref{2d-chi-H}, replace
initial conditions 
\eqref{ic<0} by the following boundary condition at infinity
\begin{gather}
\theta_{m,n}\to0\quad\mathrm{for}\quad\bb x\to\infty,
\label{2d-statics-bc-infty-nonF}
\end{gather}
and drop out the terms involving the time derivatives in 
\eqref{2D-maineq}. This yields the following equation
\begin{equation}
    \Big(\eta^2\L_{m,n} - \tfrac {c^2}4
\big(\mathscr D_{m}\partial_{x_1}+
\mathscr D_n\partial_{x_2}\big)^2  
\Big)\,
    \theta_{m,n} 
    =
    \eta\chio\de(\mathbf x)\L_{m,n}\delta_{m;0}\delta_{n;0}.
\label{2d-maineq-stat}
\end{equation}

Now we apply the discrete Fourier transform 
with respect to the variables $m,\ n$
to 
Eq.~\eqref{2d-maineq-stat}:
\begin{gather}
\F\theta{}n(p_1,p_2,\bb x,t)=\sum_{m,n=-\infty}^{\infty} \theta_{m,n}(\bb x)
\exp (-\I mp_1-\I np_2),
\label{F-theta-eta0}
\end{gather}
where subscript $F$ is the symbol of the discrete Fourier transform,
$p_1,\ p_2$ are the discrete Fourier transform parameters.
Using the shift property for the discrete Fourier transform, one gets
\begin{gather}
\big(\L_{m\,n}\theta_{m,n}\big)_F=-4\left({\sin^2\frac {p_1}2+\sin^2 \frac {p_2}2}\right)
\theta_F
,\\
\big(\mathscr D_{m}\theta_{m,n}\big)_F=2\I\,\sin{p_1}
\theta_F
,\qquad
\big(\mathscr D_{n}\theta_{m,n}\big)_F=2\I\,\sin{p_2}
\theta_F
,\\
\big((\mathscr D_{m}\partial_{x_1}+
\mathscr D_n\partial_{x_2})\theta_{m,n}\big)_F=
2\I\,({\sin p_1 \mathbf i+\sin p_2 \mathbf j})
\cdot\nabla
\theta_F
,
\\
(\delta_{m;0}\delta_{n;0})_F=1,
\end{gather}
where 
\begin{equation}
\nabla=\partial_{x_1}\bb i+\partial_{x_2}\bb j,
\end{equation}
and dot stands for the scalar product.
Multiplying the both sides of the transformed equation \eqref{2d-maineq-stat} on 
$\big(\L_{m\,n}\big)_F^{-1}$
results in\footnote{In previous paper
\cite{gavrilov2018heat} we inverse the corresponding difference operator
directly in space domain (using the identities of the calculus of finite differences), 
not in the Fourier domain.}
{the following equation}:
\begin{equation}
\eta^2\theta_F-
(\bs{\mathscr C}\cdot\nabla)^2\theta_F
=
\eta\bar\chi_0\de(\bb x)
\label{2d-statics}
\end{equation}
wherein
\begin{gather}
\bs{\mathscr C}=\breveC
({\sin p_1 \mathbf i+\sin p_2 \mathbf j})
,
\label{scrC}
\\
\breveC=\frac c
{2\sqrt{\sin^2\frac {p_1}2+\sin^2 \frac {p_2}2}}.
\label{breveC}
\end{gather}
Vector $\bs{\mathscr C}$ coincides with 
\cite{Kuzkin-Krivtsov-accepted}
the group velocity in the lattice under
condition of zero dissipation $\eta=0$.

Thus, Eq.~\eqref{2d-statics} describes 
the steady-state distribution for the Fourier images of the quantities
$\theta_{m,n}$ defined by 
Eq.~\eqref{theta_m_n} (in particular,
the kinetic temperature $T\equiv\theta_{0,0}$),
caused by a point heat source of a constant intensity.
According to conditions
\eqref{2d-statics-bc-infty-nonF},
Eq.~\eqref{2d-statics} should be supplemented with the following boundary
condition:
\begin{gather}
\theta_F\to0\quad\mathrm{for}\quad\bb x\to\infty.
\label{2d-statics-bc-infty}
\end{gather}

Equation \eqref{2d-statics} can be rewritten as 
\begin{equation}
\eta^2\theta_F-
(\bs{\mathscr C}\cdot\bs{\mathscr C})(\breve{\bs{\mathscr C}}\cdot\nabla)^2\theta_F
=\bar\chi_0\eta\delta(\bb x).
\label{2d-statics-mod}
\end{equation}
Here and in what follows,
\begin{equation}
\breve {\mathbf a}=\frac {\mathbf a}{|\mathbf a|}
\end{equation}
for any non-zero vector $\mathbf a$,
\begin{equation}
\breve{\mathbf a}_\perp=(\mathbf i\times\mathbf j)\times \breve{\mathbf a}
\end{equation}
for any unit vector $\breve{\mathbf a}$, $\times$ is the cross product.
Delta-function in the right-hand side of 
\eqref{2d-statics-mod}
can be represented as 
\begin{equation}
\delta(\bb x)=\delta_1(x_1)\,\delta_1(x_2)
=\delta_1(\bb x\cdot{\breve{\bb n}})
\,\delta_1(\bb x\cdot{\breve{\bb n}}_\perp),
\label{2d-delta-repr}
\end{equation}
where 
$\breve{\bb n}$
is an arbitrary unit vector
such that 
\begin{equation}
\breve{\bb n} \cdot (\bb i\times\bb j)=0.
\end{equation}
We introduce angular variable $\alpha=\arctan\frac{x_2}{x_1}$ such that
\begin{equation}
\bb x=|\bb x|(\cos \alpha\, \bb i+ \sin \alpha\, \bb j).
\label{2d-r}
\end{equation}
Due to symmetry, without loss of generality we can
assume that (see Figure~\ref{lattice.eps})
\begin{equation}
0\leq\alpha\leq\pi/4.
\label{alpha-restrict}
\end{equation}

The solution of 
\eqref{2d-statics-mod} satisfying
\eqref{2d-statics-bc-infty}
is
\begin{equation}
\theta_F=
\frac\chio{2|\bs{\mathscr C}|}
\exp\Big(-\frac\eta{|\bs{\mathscr C}|}\,|\bb x\cdot\breve{\bs{\mathscr C}}|\Big)
\,\delta_1(\bb x\cdot\breve{\bs{\mathscr C}}_\perp)
=
\frac\chio{2|\bb x|}
\exp\Big(-\frac\eta{|\bs{\mathscr C}|}\,|\bb x\cdot\breve{\bs{\mathscr C}}|\Big)
\,\delta_1(\breve{\bb x}_\perp\cdot{\bs{\mathscr C}}).
\end{equation}
Here we have used 
Eq.~\eqref{2d-delta-repr} wherein $\breve {\bb n}=\breve {\bs{\mathscr C}}$, and the
following formulas:
\begin{gather}	
\delta_1(Ay)=\frac{1}{|A|}\delta_1(y),\qquad A\in\mathbb R,\ A\neq0;\\
\bb x\cdot\breve{\bs{\mathscr C}}_\perp=0
\quad\Longleftrightarrow\quad
\breve{\bb x}_\perp\cdot{\bs{\mathscr C}}=0.
\end{gather}

Applying the inverse Fourier transform, one get the expression for the
kinetic temperature $T=\theta_{0,0}$:
\begin{multline}
\kappaT=\frac 1{4\pi^2}\iint_{-\pi}^\pi \F\theta \qy
n\exp(\I mp_1+\I np_2)\,\d y\,\bigg|_{m=n=0}\\=
\frac 1{4\pi^2}\,
\frac \chio{2|\bb x|}
\iint_{-\pi}^{\pi}
\exp
\Big(-\frac\eta{|\bs{\mathscr C}(p_1,p_2)|}\,|\bb x\cdot\breve{\bs{\mathscr C}}|\Big)
\,\delta_1\big(\breve{\bb x}_\perp\cdot{\bs{\mathscr C}(p_1,p_2)}\big)
\,\d p_1\, \d p_2.
\label{2d-maineq-prealpha}
\end{multline}

Since the integrand in the right-hand side of Eq.~\eqref{2d-maineq-prealpha} 
contains the Dirac delta-function $\delta_1$ in one-dimensional space, 
formula \eqref{2d-maineq-prealpha} can be
rewritten in a simpler form involving a single integral
(see~Appendix~\ref{2d-app-A}):
\begin{equation}
\kappaT=
\kappaT_1+\kappaT_2\equiv
\frac \chio{4\pi^2|x_1|}
\sum_{j=1}^2
\int_{0}^{\pi}
\frac
{\exp
\left(
-\eta \left|\frac{x_1}{\sin p_1\,{\breveC\big(p_1,p_2^{(j)}\big)}}\right|
\right)
}
{\breveC
\big(p_1,p_2^{(j)}\big)
\sqrt{1-\tan^2\alpha\,\sin^2p_1}
}
\,\d p_1,
\label{2dkappa-final}
\end{equation}
where $\breveC
\big(p_1,p_2^{(j)}\big)$ $(j=1,2)$ are defined by Eqs.~\eqref{2d-breve-C1},
\eqref{2d-breve-C2}. If necessary, the far-field asymptotics of the solution 
\eqref{2dkappa-final} can be calculated by means of the Laplace method
\cite{Fedoruk-Saddle}. 

\subsection{The case $\alpha=0$}

In the case $\alpha=0$,  
i.e.\ for two orthogonal rows of the
lattice that contain the heat source located at 
$\bb x=\mathbf 0$
(see Figure~\ref{lattice.eps}),
it is possible to obtain the solution in a simpler form. This is done in
Appendix~\ref{2d-app-B}. The final
result is $T=T_1+T_2$, where $T_1$ and $T_2$ are defined by Eqs.~
\eqref{2d-a0-kappa1},
\eqref{2d-a0-kappa2}, respectively.

\subsection{The case $\alpha=\pi/4$}
\label{2d-sec-singular}
Consider the particular case $\alpha=\pi/4$,
i.e.\ the main diagonals of the lattice with respect to a point heat 
source position
(see Figure~\ref{lattice.eps}).
One has
\begin{gather}	
\left.\frac1{\sqrt{1-\tan^2\alpha\sin^2p_1}}\right|_{\alpha=\frac\pi4}=
\frac1{|\cos p_1|}\xrightarrow[p_1\to\pm\frac\pi2]{}+\infty,
\\
\breveC
\big(p_1,p_2^{(j)}\big)\Big|_{\alpha=\frac\pi4,\ p_1=\pm\frac\pi2}=\frac12
\qquad(j=1,2).
\end{gather}
Thus, the integrands in the right-hand side of 
\eqref{2dkappa-final} have non-integrable singularities at $p_1=\pm\pi/2$, and
therefore
$\kappaT=+\infty$. In presence of the dissipation, this seems to be a quite
unexpected result, which is discussed in Section~\ref{2d-sec-numerics}.

\subsection{The solution for a row {$x_2/a=j$}}
Consider now the distribution of the kinetic temperature over a row of the
lattice
{$x_2/a=j\in\mathbb Z$}.
Provided that assumption
\eqref{alpha-restrict} is true,
the analytic expression for the stationary distribution of the kinetic temperature 
$\kappaT$ is
\eqref{2dkappa-final}, where 
\begin{equation}
\tan\alpha=\frac{|x_2|}{|x_1|},\qquad |x_2|<|x_1|.
\end{equation}
For $\alpha>\pi/4$, due to symmetry with respect to the main diagonal, one has 
\begin{equation}
\kappaT=
\frac \chio{4\pi^2|x_2|}
\sum_{j=1}^2
\int_{0}^{\pi}
\frac
{\exp
\left(
-\eta \left|\frac{x_2}{\sin p_1\,{\breveC\big(p_1,p_2^{(j)}\big)}}\right|
\right)
}
{\breveC
\big(p_1,p_2^{(j)}\big)
\sqrt{1-\tan^2\alpha\,\sin^2p_1}
}
\,\d p_1,
\label{2dkappa-final-sym}
\end{equation}
where 
\begin{equation}
\tan\alpha=\frac{|x_1|}{|x_2|},\qquad |x_1|<|x_2|.
\end{equation}

\section{Numerics}
\label{2d-sec-numerics}
In this section, we present the results of the numerical solution of the system of
stochastic differential equations \eqref{2D-1}--\eqref{2D-Winer} 
with initial conditions
\eqref{2D-ic-stochastic}. It is useful to rewrite Eqs.~\eqref{2D-1}--\eqref{2D-Winer}
in the dimensionless form
\begin{equation}
\begin{aligned}	
&\d {\tilde{v}}_{i,j} =  (\L_{i,j} {\tilde{u}}_{i,j} - \eta {\tilde{v}}_{i,j})
\,\d \tilde t +
\tilde b_{i,j} \rho_{i,j} \sqrt{\d \tilde t},\\ 
&\d {\tilde{u}}_{i,j}= {\tilde{v}}_{i,j}\, \d\tilde t, 
\end{aligned}
\label{2D-dimless}
\end{equation}
where 
\begin{equation}
\tilde u\=\frac u a,\quad \tilde v\=\frac v c,\quad \tilde t\={\omega_0} t,\quad  
\tilde b\=\frac b{c\sqrt{\omega_0}}, \quad \tilde\eta\=\frac\eta{\omega_0}.
\label{2D-all-dless}
\end{equation}
We consider the square lattice of $(2N+1)^2$ particles with the following 
boundary conditions 
\begin{equation}
\begin{aligned}
    u_{-N,i} &= u_{-N+1,i},  &\qquad &u_{N,i} = u_{N-1,i},
    \\
    v_{-N,i} &= v_{-N+1,i},  &\qquad &v_{N,i} = v_{N-1,i},
    \\
    u_{i,-N} &= u_{i,-N+1},  &\qquad &u_{i,N} = u_{i,N-1},
    \\
    v_{i,-N} &= v_{i,-N+1},  &\qquad &v_{i,N} = v_{i,N-1},
\end{aligned}
\end{equation}
where $i=\overline{-N, N}$.
Actually, the specific form of this boundary conditions 
is not very important in our calculations, since
we take large enough $N$ such that the wave reflections from 
the boundaries do not occur.\footnote{In \cite{Kuzkin-Krivtsov-accepted} it is shown that for the
non-dissipative system the continual non-stationary solution caused by a point pulse source
is non-zero only in the circle with radius $ct$. It may be shown that the
analogous result is true also for the dissipative system.}
To obtain a numerical solution in the case of the point source of the heat
supply located at $i=0$, $j=0$, we assume that 
$\tilde b_{i,j}\rho_{i,j}=\delta_{i;0}\delta_{j;0}\tilde b\rho_{i,j}$ and
use the scheme 
\begin{equation}
\begin{aligned}	
\Delta {\tilde{v}}_{i,j}^k &=  (\L_{{i,j}} {\tilde{u}}_{i,j}^k - \eta {\tilde{v}}_{i,j}^k)
\Delta {\tilde{t}} +
\tilde b\delta_{i;0}\delta_{j;0} \rho^k_{i,j} \sqrt{\Delta {\tilde{t}}},\\ 
\Delta {\tilde{u}}_{i,j}^k &= {\tilde{v}}_{i,j}^{k+1} \Delta {\tilde{t}},
\\
\tilde v^{k+1}_{{i,j}} &= \tilde v^{k}_{{i,j}} + \Delta \tilde v^{k}_{{i,j}}
,
\\
\tilde u^{k+1}_{{i,j}} &= \tilde u^{k}_{{i,j}} + \Delta \tilde u^{k}_{{i,j}},
\end{aligned}
\label{2D-scheme}
\end{equation}
where  $i,j=\overline{-N, N}$. 
Here the symbols  with
superscript $k$ denote the corresponding quantities at $\tilde
t=\tilde t^k\=k\Delta \tilde t$:
${\tilde{u}}_{i,j}^k={\tilde{u}}_{i,j}(\tilde t^k),\ \tilde v_{i,j}^k=\tilde v_{i,j}(\tilde t^k)$;
$\rho^k$ are normal random numbers that satisfy \eqref{2D-82} generated for all
$\tilde t^k$.  Without loss of generality we can take $\tilde b=1$. Due to
symmetry of the problem we can perform the calculations only for $1/8$ of the whole
lattice ($i=\overline{0,N}$, $j=\overline{0,i}$).

We perform a series of $r=1\dots R$ realizations of these calculations (with
various independent 
$\rho^k_{(r)}$) and get the corresponding particle velocities $\tilde v^k_{(r)i,j}$. In
accordance with
\eqref{2D-temp-def},      
{in order to
obtain the dimensionless kinetic temperature 
\begin{equation}
\tilde T=\frac
{Tk_B}{mc^2},	
\label{2D-kin-temp-dless}
\end{equation}
we should average 
the doubled dimensionless kinetic energies}:
\begin{equation}
\tilde T_{i,j}^k =\frac{1}{R}\sum_{r=1}^{R} (\tilde v^k_{(r)i,j})^2.
\label{2D-kin-temp-num}
\end{equation}

Numerical results  
\eqref{2D-kin-temp-num}
for the kinetic temperature
can be compared with the analytical 
steady-state solution
\eqref{2dkappa-final},
expressed in the dimensionless form wherein
$a=1,\ c=1$, $x_1=i$, $x_2=j$, and
\begin{equation}
\chio=\frac{\tilde b^2}2.
\end{equation}
Note that the factor $1/2$ in the right-hand sides of the last formula
appears according to Eq.~\eqref{2D-chi-def}.

\begin{figure}[htbp]	
\centering\includegraphics[width=0.85\columnwidth]{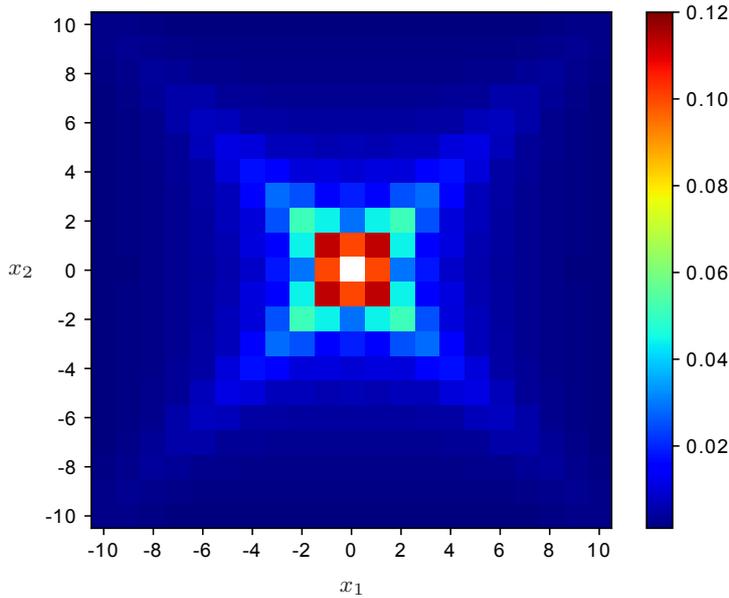}
\caption{The central zone of the kinetic temperature distribution pattern in a square
two-dimensional lattice, caused by a
constant point source (the value at the central point $\tilde
T_{0,0}\simeq0.79$ is shown by the white color)}
\label{2d-general.eps}
\end{figure}

\begin{figure}[htbp]	
\centering\includegraphics[width=0.85\columnwidth]{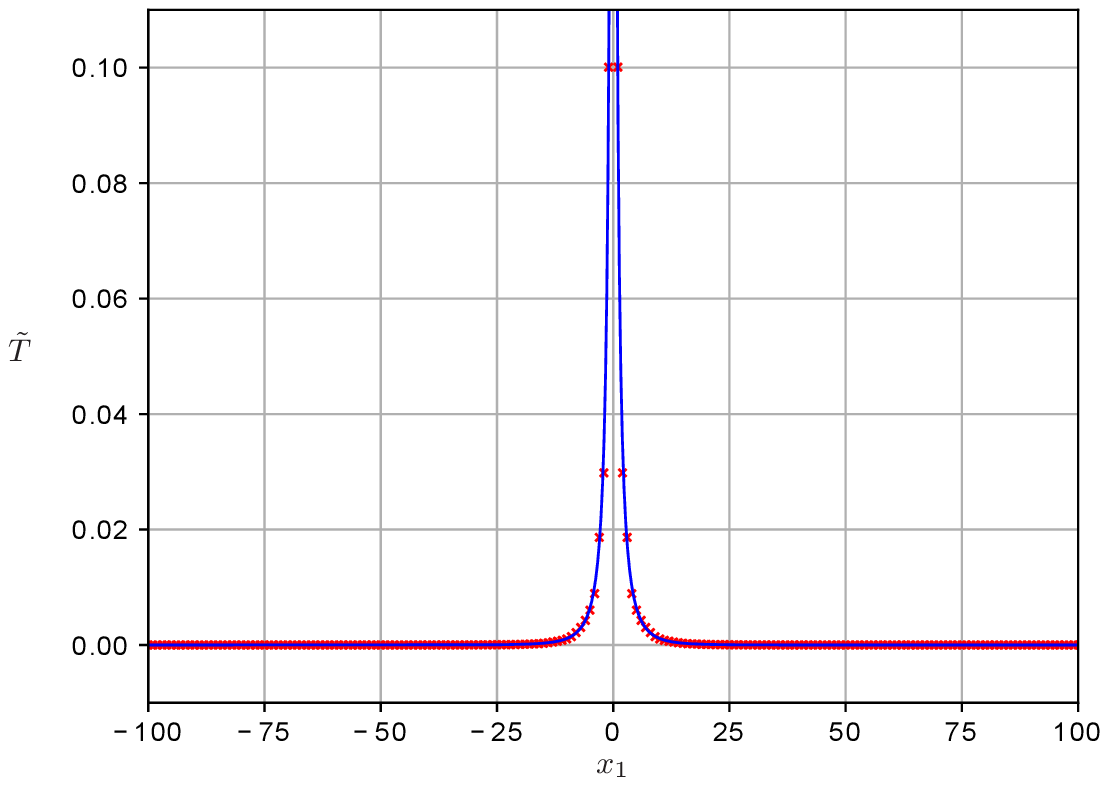}
\caption{Comparing the steady-state analytical solution in the form of Eqs.~\eqref{2d-a0-kappa1},
\eqref{2d-a0-kappa2}
for the row $x_2=0$
(the blue solid line) and the
numerical solution (the red crosses)}
\label{2d-a0.eps}
\end{figure}

\begin{figure}[htbp]	
\centering\includegraphics[width=0.85\columnwidth]{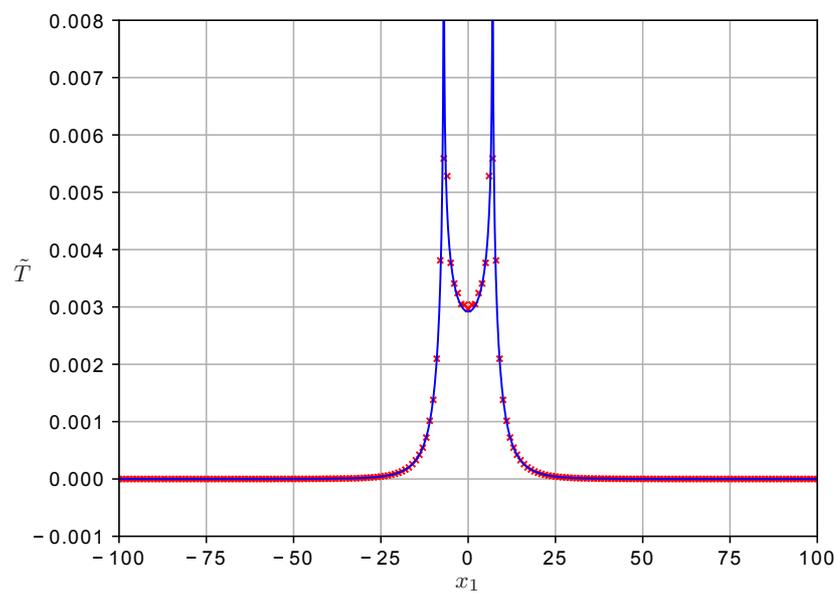}%
\caption{Comparing the steady-state analytical solution 
in the form of Eqs.~\eqref{2dkappa-final}, \eqref{2dkappa-final-sym}
for the row $x_2=7$
(the blue solid line) and the
numerical solution (the red crosses)
}
\label{2d-7raw.eps}
\end{figure}

\begin{figure}[htbp]	
\centering{\includegraphics[width=0.85\columnwidth]%
{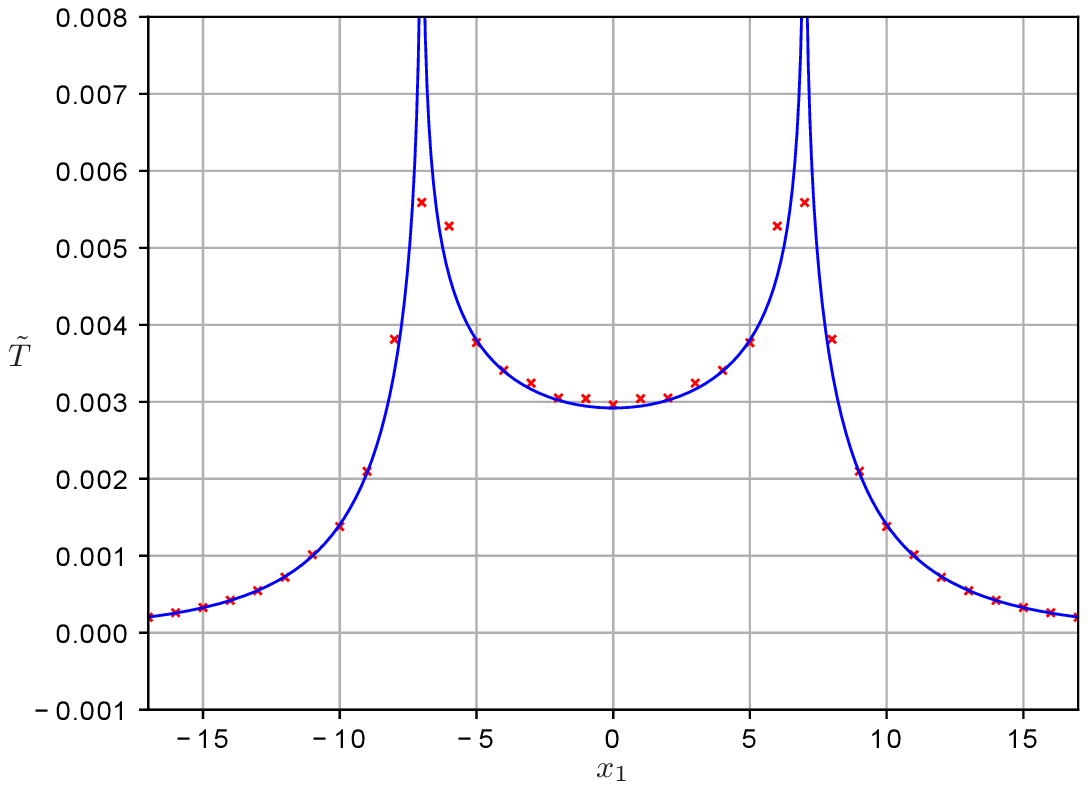}
}
\caption{Comparing the steady-state analytical solution 
in the form of Eqs.~\eqref{2dkappa-final}, \eqref{2dkappa-final-sym}
for the central zone of row $x_2=7$
(the blue solid line) and the
numerical solution (the red crosses)
}
\label{2d-7raw-c.eps}
\end{figure}

\begin{figure}[htbp]	
\centering{\includegraphics[width=0.85\columnwidth]%
{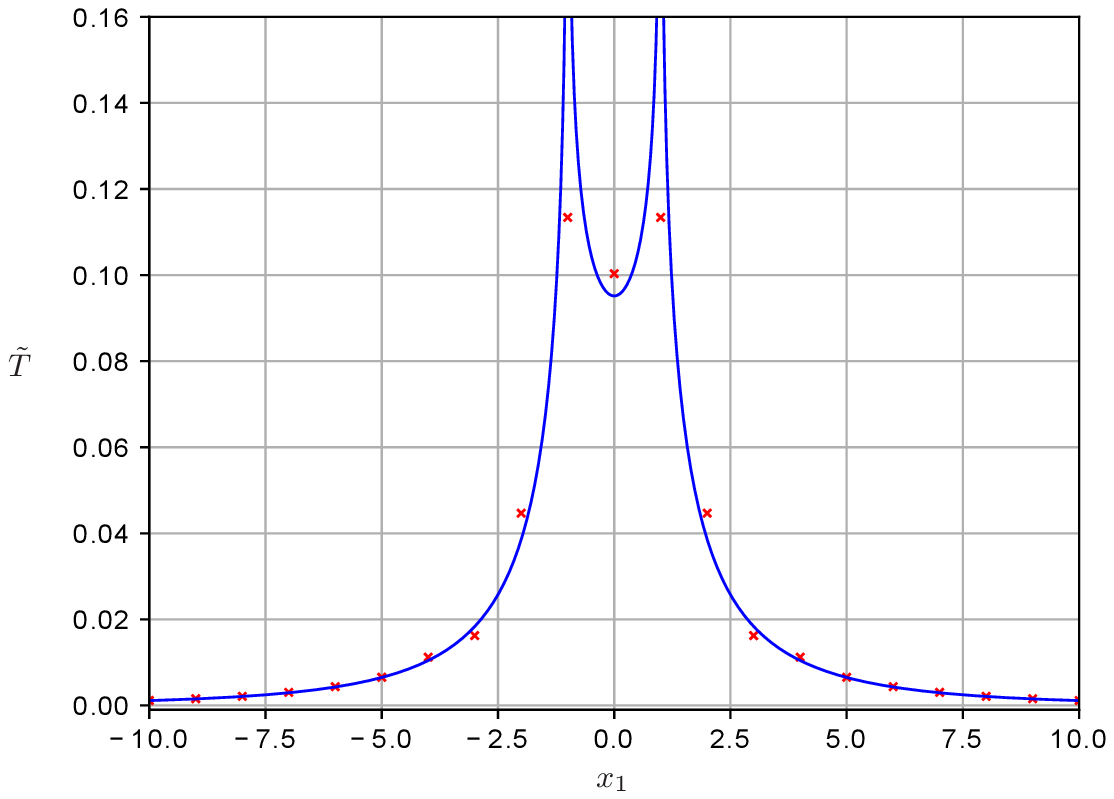}
}
\caption{Comparing the steady-state analytical solution 
in the form of Eqs.~\eqref{2dkappa-final}, \eqref{2dkappa-final-sym}
for the central zone of row $x_2=1$
(the blue solid line) and the
numerical solution (the red crosses)
}
\label{2d-1raw-c.eps}
\end{figure}

The comparison between analytical 
and numerical results is presented in
Figures~\ref{2d-general.eps}--\ref{2d-1raw-c.eps}.
All calculations were performed for the following values of the problem
parameters: $\tilde\eta=0.1$, $N=100$,
$\tilde t=100$, $R=100000$. The time step is $\Delta{\tilde t}=0.025$.
To perform the numerical
calculations we use 
{\sc SciPy} software \cite{scipy}.

In Figure~\ref{2d-general.eps} one can see the general view of a central zone of
the kinetic temperature 
distribution pattern in the lattice. One can observe that most of the heat
propagates along the main diagonals of the lattice (with respect to a point
source position). This result is in a qualitative agreement with results obtained earlier
in studies 
\cite{mielke2006macroscopic,harris2008energy}, where energy transport in
lattices are discussed in the deterministic case, and also with results of
\cite{Kuzkin-Krivtsov-accepted}.

In Figure~\ref{2d-a0.eps} we compare the steady-state analytical solution 
in the form of Eqs.~\eqref{2d-a0-kappa1},
\eqref{2d-a0-kappa2}
for the row $x_2=0$
(the blue solid line) and the
numerical solution (the red crosses).

In Figure~\ref{2d-7raw.eps} we compare the steady-state analytical solution 
in the form of Eqs.~\eqref{2dkappa-final}, \eqref{2dkappa-final-sym}
for the row $x_2=7$
(the blue solid line) 
and the
numerical solution (the red crosses). The corresponding view of a central zone
of the kinetic temperature distribution 
is given in Figure~\ref{2d-7raw-c.eps}.
One can see that numerical value at the point $|x_1|=7$ of the
lattice at the main diagonal is finite, whereas the analytic continuum solution 
is singular at this point. 

Finally, in Figure~\ref{2d-1raw-c.eps} we compare the steady-state analytical solution 
in the form of Eqs.~\eqref{2dkappa-final}, \eqref{2dkappa-final-sym}
for the row $x_2=1$
(the blue solid line) 
and the
numerical solution (the red crosses) in a central zone near the source.

\begin{figure}[htbp]	
\centering{\includegraphics[width=0.85\columnwidth]%
{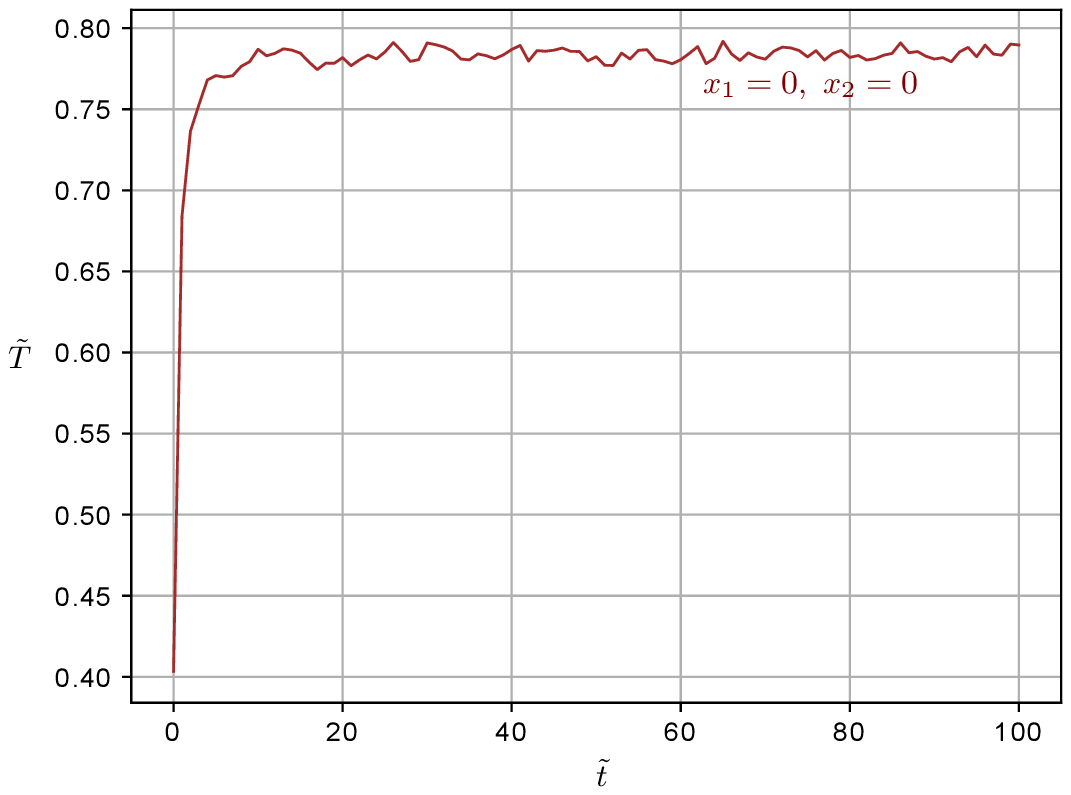}
}\\%
\centering{\includegraphics[width=0.85\columnwidth]{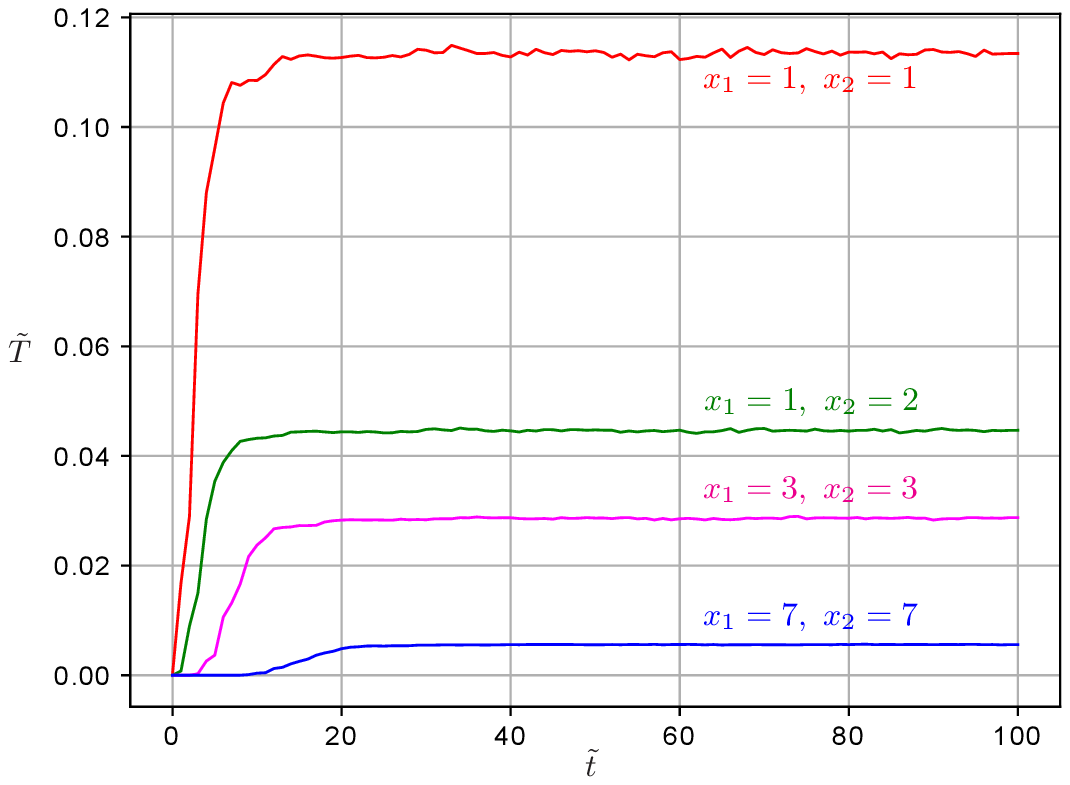}%
}
\caption{Plots of the numerical solution for the 
kinetic temperature at several fixed
positions
versus the time (the plots are drawn with time step $1.0$) 
}
\label{time.eps}
\end{figure}

One can see that in all cases, the analytical and
numerical solutions are in a very good agreement everywhere except the main
diagonals of the lattice $|x_1|=|x_2|$
(see Figures~\ref{2d-a0.eps}--\ref{2d-1raw-c.eps}).
The continuous analytical
solution predicts singularities in the stationary solution at the main
diagonals (see Section~\ref{2d-sec-singular}). Usually, such a result means
that a corresponding non-stationary solution grows with time, and a 
steady-state solution at singular points does not exist. However, the numerical
calculations based on the original infinite system of stochastic ODE
contradict with such a hypothesis and
predict that 
despite 
most of the heat
propagates along the main diagonals of the lattice (see 
Figure~\ref{2d-general.eps}),
the kinetic
temperature at the main diagonals apparently converges to finite values (see 
Figure~\ref{time.eps}, where plots of the numerical solution for the 
kinetic temperature at several fixed
positions
versus the time are given).
In particular, this is true for the point
$x_1=x_2=0$ where the heat source is located (see 
Figure~\ref{time.eps}). 
Note that in one-dimensional case considered in \cite{gavrilov2018heat}
we also observe the
singularity in the continuous solution at the point where heat source is
applied, whereas the numerics predicts the finite value of the kinetic
temperature at that point\footnote{This is not discussed in
\cite{gavrilov2018heat}}. In our opinion,
such paradoxical result is caused by the continualization procedure (in
particular, by the choice of heat supply intensity in singular form
\eqref{2d-chi-H}),
and it
definitely needs an additional investigation. 
Note that in the conservative non-stochastic case the singularities at main
diagonals were discovered in
\cite{mielke2006macroscopic,harris2008energy,giannoulis2006continuum}, where
they are associated with non-smoothness of the dispersion relation.

\section{Conclusion}
\label{2d-Sec-conclusion}

In the paper, we started with equations \eqref{2D-1} for stochastic dynamics
of a two-dimensional harmonic square scalar lattice in a viscous environment.
We introduced in the standard way the kinetic temperature in the lattice as a
quantity proportional to the variance of the particle
velocities. The most important results of the paper are the
differential-difference equation \eqref{2D-maineq} describing non-stationary
heat propagation in the lattice and the analytical formula in the integral
form \eqref{2dkappa-final} describing the steady-state kinetic temperature
distribution in the lattice caused by a point heat source of a constant intensity.

The comparison between numerical solution of equations \eqref{2D-1} and
analytic steady-state solution \eqref{2dkappa-final} of
differential-difference equation \eqref{2D-maineq} demonstrates a a very good
agreement everywhere except the main diagonals of the lattice with respect to
the point source position
(see Figures~\ref{2d-a0.eps}--\ref{2d-1raw-c.eps}).
The
continuous analytical solution predicts singularities in the stationary
solution at the main diagonals (see Section~\ref{2d-sec-singular}), whereas
the numerical solution predicts that despite most of the heat propagates along
the main diagonals of the lattice (see Figure~\ref{2d-general.eps}), the
kinetic temperature at the main diagonals apparently converges to finite
values (see 
Figure~\ref{time.eps}).  In our opinion, such a paradoxical result can be caused
by the continualization procedure (in particular, by the choice of heat supply
intensity in singular form \eqref{2d-chi-H}).
Another possible oversimplification in physical modeling that makes the main
diagonals to be preferable directions for heat propagation, apparently,
is the choice of the potential of interaction involving for a given particle only 
four neighbours (as it is stated by Eqs.~\eqref{2D-Li}, \eqref{2D-Lij}).
This situation needs an additional investigation. 

We expect that the results obtained in the paper can be used to describe the
heat transfer in low-dimensional nanostructures and ultra-pure materials
\cite{chang2008breakdown,xu2014length,goldstein2007mechanics}. 
The subject of our future work is to generalize the theoretical results
to the case of the graphene lattice and to verify the results by means of 
experiments with the laser heating of graphene 
\cite{hwang2016measuring,indeitsev2017two}.

\section*{Acknowledgements} The authors are grateful to V.A.~Kuzkin, A.S.~Murachev, 
and E.V.~Shishkina for useful and
stimulating discussions.

\appendix
\section{Calculation of the steady-state solution in the integral form}
\label{2d-app-A}

To calculate the right-hand side of Eq.~\eqref{2d-maineq-prealpha}, one needs to use
the formula (see \cite{G-Sh-1})
\begin{equation}
\int_I \delta_1\big(f(\qy)\big)\,\d\qy=\sum_j \frac{1}{|f'(\qy_j)|}.
\label{int-delta}
\end{equation}
where $y_j$ are the roots of $f (y)$ lying inside the interval $I$.

\begin{figure}[htbp]	
\centering\includegraphics[width=0.65\textwidth]{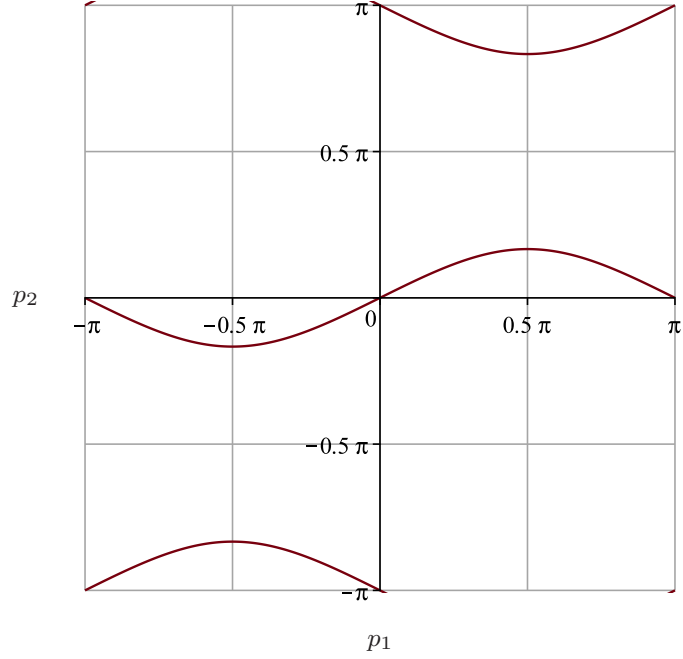}
\caption{Structure of roots $p_2^{(j,k)}$ defined by 
\eqref{2d-p12}
such that $p_2^{(j,k)}\in[-\pi,\pi]$ in the case $0<\alpha<\pi/2$ (here
$\alpha=\arctan{\tfrac12})$}
\label{p-roots.eps}
\end{figure}

One has 
\begin{gather}
\bb x_\perp=|\bb x|(\cos \alpha\, \bb j- \sin \alpha\, \bb i),\\
\breve{\bb x}_\perp\cdot\bC
=\breveC(\cos\alpha\sin p_2-\sin\alpha\sin p_1),
\end{gather}%
\begin{multline}
\breve{\bb x}_\perp\cdot\bC=0
\quad\Longleftrightarrow\quad
\cos\alpha\sin p_2-\sin\alpha\sin p_1=0
\quad\Longleftrightarrow\quad
\frac{\sin p_2}{\sin p_1}=\tan\alpha\\
\Longleftrightarrow\quad
p_2=p_2^{(j,k)},\quad j=1,2;\ k\in\mathbb Z;
\end{multline}
where
\begin{equation}
\begin{gathered}
p_2^{(1,k)}\equiv\arcsin(\tan \alpha\sin p_1)+2\pi k
,
\\
p_2^{(2,k)}\equiv\pi-\arcsin(\tan \alpha\sin p_1)+2\pi k
,
\end{gathered}
\label{2d-p12}
\end{equation}
such that $p_2^{(j,k)}\in[-\pi,\pi]$.
The typical structure of roots $p_2^{(j,k)}$ in the case $0<\alpha<\pi/4$ is
presented in Figure~\ref{p-roots.eps}.
For $p_1\neq0$ there exist exactly two roots lying in the interval $[-\pi,\pi]$.
The first one corresponds to the choice $j=1,\ k=0$, and the second one
corresponds to the choice $j=2,\ k=0$ for $p_1>0$ and $j=2,\ k=-1$ for
$p_1<0$.
Thus, for $0<\alpha<\pi/4$ we put
\begin{equation}
\begin{gathered}
p_2^{(1)}\equiv\arcsin(\tan \alpha\sin p_1)
,
\\
p_2^{(2)}\equiv\pm\pi-\arcsin(\tan \alpha\sin p_1)
,
\quad \pm p_1>0.
\end{gathered}
\label{2d-p12-simple}
\end{equation}

Applying now formula
\eqref{int-delta}, one gets
\begin{gather}
\delta_1(
\breve{\bb x}_\perp\cdot\bC
)
=
\sum_{j=1}^2
\frac{\delta_1\big(p_2-p_2^{(j)}\big)}{\breveC|\cos\alpha|
\,\big|\cos p_2^{(j)}\big|},\\
\kappaT=\sum_j \kappaT_j\equiv
\frac \chio{8\pi^2 |\bb x|\,|\cos\alpha|}
\sum_{j=1}^2
\int_{-\pi}^{\pi}
\frac1
{\breveC\big|\cos p_2^{(j)}
\big|}
\,
{\exp
\left(-\frac{\eta\,|\bb x\cdot\breve{\bs{\mathscr C}}|}
{\big|\bs{\mathscr C}
\big|}\right)
}
\,\d p_1.
\label{ppreT}
\end{gather}
According to 
\eqref{2d-p12} one has:
\begin{gather}
\cos \arcsin p=\sqrt{1-p^2},\\
|\cos p_2^{(j)}|
=\sqrt{1-\tan^2\alpha\,\sin^2p_1}
,\quad
j=1,2.
\end{gather}
Using 
\eqref{breveC}, and taking into account identities 
\begin{gather}
\sin^2 \frac p2=\frac{1-\cos p}2,
\\
\cos^2\alpha=\frac{1}{1+\tan^2\alpha},
\end{gather}
one obtains
\begin{gather}
\breveC\big(p_1,p_2^{(1)}\big)=
\frac c
{2\sqrt{\sin^2\frac {p_1}2+\frac{1-\cos\arcsin(\tan \alpha\sin p_1)}2  }}
=
\frac c
{2\sqrt{\sin^2\frac {p_1}2+\frac{1-
\sqrt{1-\tan^2\alpha\,\sin^2p_1}}2  }},
\label{2d-breve-C1}
\\
\breveC\big(p_1,p_2^{(2)}\big)=
\frac c
{2\sqrt{\sin^2\frac {p_1}2+\frac{1+\cos\arcsin(\tan \alpha\sin p_1)}2  }}
=
\frac c
{2\sqrt{\sin^2\frac {p_1}2+\frac{1+\sqrt{1-\tan^2\alpha\,\sin^2p_1}}2  }}.
\label{2d-breve-C2}
\end{gather}
According to 
\eqref{r-def},
\eqref{scrC}, one gets
\begin{gather}
\bs{\mathscr C}\big(p_1,p_2^{(j)}\big)
=\breveC\big(p_1,p_2^{(j)}\big)\sin p_1 
({\mathbf i+\tan \alpha\, \mathbf j}),
\\
\Big|\bs{\mathscr C}\big(p_1,p_2^{(j)}\big)\Big|=
\breveC\big(p_1,p_2^{(j)}\big)|\sin p_1|
\sqrt{1+\tan^2\alpha},
\\
\breve{\bs{\mathscr C}}\big(p_1,p_2^{(j)}\big)=
\frac
{\sign p_1}
{\sqrt{1+\tan^2\alpha}}
\,
{(\mathbf i+\tan \alpha\, \mathbf j)}
,
\\
\bb x=x_1(\bb i+\tan\alpha\,\bb j),
\\
\bb x\cdot\breve{\bs{\mathscr C}}\big(p_1,p_2^{(j)}\big)=
{x_1\sign p_1}
{\sqrt{1+\tan^2\alpha}}
,
\\
\frac{\Big|\bb x\cdot\breve{\bs{\mathscr C}}\big(p_1,p_2^{(j)}\big)\Big|}
{\Big|\bs{\mathscr C}\big(p_1,p_2^{(j)}\big)\Big|}=
\left|\frac{x_1}{\sin p_1\,{\breveC\big(p_1,p_2^{(j)}\big)}}\right|,
\\
|\bb x|=\sqrt{x_1^2+x_2^2}
=
|x_1|\sqrt{1+\tan^2\alpha},
\\
|\bb x|\,|\cos\alpha|=
|x_1|,
\end{gather}
where $j=1,2$.
Finally, substituting of these expressions into 
Eq.~\eqref{ppreT} and simplifying of expression obtained, results in the formula
for the solution in the solution in the integral form:
\begin{equation}
\kappaT=
\kappaT_1+\kappaT_2=
\frac \chio{4\pi^2|x_1|}
\sum_{j=1}^2
\int_{0}^{\pi}
\frac
{\exp
\left(
-\eta \left|\frac{x_1}{\sin p_1\,{\breveC\big(p_1,p_2^{(j)}\big)}}\right|
\right)
}
{\breveC
\big(p_1,p_2^{(j)}\big)
\sqrt{1-\tan^2\alpha\,\sin^2p_1}
}
\,\d p_1,
\label{2dkappa-final-app}
\end{equation}

\section{Calculation of the steady-state solution in the particular case
$\alpha=0$}
\label{2d-app-B}

In the case under consideration, due to 
\eqref{2d-p12}
one has
\begin{equation}
\begin{gathered}
p_2^{(1,0)}=0
,
\\
p_2^{(2,0)}=\pi,\qquad
p_2^{(2,-1)}=-\pi.
\end{gathered}
\end{equation}
Since in the case $p=p_2$ the roots lie at the boundaries of
the interval $[-\pi,\pi]$,  we take the corresponding contributions multiplied by $1/2$.
Since these contributions are equal to each other, we can put
\begin{equation}
\begin{gathered}
p_2^{(1)}=0
,
\\
p_2^{(2)}=\pi
\end{gathered}
\end{equation}
and use the formulas obtained above.
According to Eqs.~\eqref{2d-breve-C1},
\eqref{2d-breve-C2} one gets 
\begin{gather}
\breveC\big(p_1,p_2^{(1)}\big)=
\frac c
{2|\sin\frac {p_1}2|
},
\label{breveC0}
\\
\breveC\big(p_1,p_2^{(2)}\big)=
\frac c
{2\sqrt{\sin^2\frac {p_1}2+1}
},
\label{breveC-pi}
\end{gather}
Thus,
\begin{equation}
\kappaT_1=
\frac \chio{4\pi^2 {|x_1|}}\int_0^\pi\frac2c
\left|
\sin\frac {p_1}2
\right|
\,
{\exp
\left(-\frac{\eta{|x_1|}}
{c\left|\cos\frac {p_1}2\right|
}\right)
}
\,\d p_1
=
\frac \chio{\pi^2 {|x_1|}c}\int_{0}^{\pi/2}
\sin y
\,
{\exp
\left(-\frac{\eta{|x_1|}}
{c\cos y
}\right)
}
\,\d y
.
\end{equation}
Let $\gamma=\frac1{\cos y}$. One has
\begin{gather}
\d \gamma=\frac {\sin y}{\cos^2 y}\,\d y
=\gamma^2\sin y\,\d y,
\end{gather}
and, therefore, 
\begin{equation}
\kappaT_1=
\frac \chio{\pi^2 {|x_1|}c}\int_1^\infty
\gamma^{-2}
\,
{\exp
\left(-\frac{\eta{|x_1|}\gamma}
{c}\right)
}
\,\d\gamma
=
\frac \chio{\pi^2 {|x_1|}c}
\left(
\exp
\left(
-\frac{\eta{|x_1|}}{c}
\right)
-\frac{\eta{|x_1|}}{c}
\Ei_1
\left(
\frac{\eta{|x_1|}}{c}
\right)
\right)
,
\label{2d-a0-kappa1}
\end{equation}
where
\begin{equation}
\Ei_1(z)=\int_1^\infty\frac{\exp(-\gamma z)}\gamma\,\d\gamma,\quad\Re (z)>0
\end{equation}
is the exponential integral \cite{abramowitz1972handbook}.
For $T_2$ one gets  the solution in the integral form
\begin{equation}
\kappaT_2=
\frac \chio{2\pi^2 {|x_1|}c}\int_{0}^\pi
\sqrt{\sin^2\frac {p_1}2+1}
\,\,
{\exp
\left(-\frac{2\eta{|x_1|}\sqrt{\sin^2\frac {p_1}2+1}}
{c\sin{p_1}
}\right)
}
\,\d p_1.
\label{2d-a0-kappa2}
\end{equation}

\selectlanguage{english}


\end{document}